\documentclass[showpacs,preprintnumbers,amsmath,amssymb,superscriptaddress,aip]{revtex4-1}
\usepackage{graphicx}

\def\beq{\begin{equation}}
\def\eeq{\end{equation}}
\def\beqar{\begin{eqnarray}}
\def\eeqar{\end{eqnarray}}

\def\para{\parallel}

\newcommand{\pdiff}[2]{\frac{\partial#1}{\partial#2}}

\newcommand{\pdt}{\partial_t}
\newcommand{\pdr}{\partial_r}

\newcommand{\pdrr}{\partial^2_r}

\def\grad{\nabla}

\newcommand{\gradpar}{\grad_\parallel}
\newcommand{\gradperp}{\grad_\perp}

\newcommand{\vpe}{v_{\parallel e}}

\newcommand{\nue}{\nu_{e}}

\newcommand{\nuin}{\nu_{in}}
\newcommand{\kpe}{\kappa_{\parallel e}}

\newcommand{\fmie}{\frac{m_i}{m_e}}
\newcommand{\fmei}{\frac{m_e}{m_i}}

\begin{document}

\title{Energy dynamics in a simulation of LAPD turbulence}

\author{B. Friedman}
\email{friedman@physics.ucla.edu}

\author{T.A. Carter}

\affiliation{Department of Physics and Astronomy, University of California, Los Angeles, California 90095-1547, USA}

\author{M.V. Umansky}
\affiliation{Lawrence Livermore National Laboratory, Livermore, California 94550, USA}

\author{D. Schaffner}

\affiliation{Department of Physics and Astronomy, University of California, Los Angeles, California 90095-1547, USA}

\author{B. Dudson}

\affiliation{Department of Physics, University of York, Heslington, York YO10 5DD, United Kingdom}

\begin{abstract}
Energy dynamics calculations in a 3D fluid simulation of drift wave turbulence in the linear Large Plasma Device (LAPD) [W. Gekelman \emph{et al.}, Rev. Sci. Inst. {\bf 62}, 2875 (1991)]
illuminate processes that drive and dissipate the turbulence.
These calculations reveal that a nonlinear instability dominates the injection of energy into the turbulence by overtaking the linear drift
wave instability that dominates when fluctuations about the equilibrium are small.
The nonlinear instability drives flute-like ($k_\parallel = 0$) density fluctuations using free energy from the background density gradient. 
Through nonlinear axial wavenumber transfer to $k_\parallel \ne 0$ fluctuations, the nonlinear instability accesses the adiabatic response, 
which provides the requisite energy transfer channel from density to potential fluctuations as well as the
phase shift that causes instability. The turbulence characteristics in the simulations agree remarkably well with experiment. When the nonlinear instability is artificially 
removed from the system through suppressing $k_\parallel=0$ modes, the turbulence develops a coherent frequency spectrum which is inconsistent
with experimental data. This indicates the importance of the nonlinear instability in producing experimentally consistent turbulence.
\end{abstract}

\maketitle

\section{Introduction}
It is common practice to study a system's linear stability properties to gain insight into turbulent dynamics. It is often easier to calculate and analyze linear modes and growth rates than
to simulate and analyze nonlinear turbulence. However, there are several situations in which linear properties can be misleading in understanding turbulent systems. First,
linear studies of magnetically confined plasmas that neglect stable branches of the linear dispersion relation often miss details of nonlinear dynamics. For example, stable eigenmodes can often
impact nonlinear dynamics by providing energy sinks and sometimes energy sources not found on the most unstable linear 
branch~\cite{baver2002,terry2002,terry2006a,terry2006b,gatto2006,terry2009,hatch2009,kim2010,makwana2011,hatch2011}. Stable eigenmodes can shift
the energy injection and dissipation ranges, making the turbulent dynamics very different from the Kolmogorov picture of hydrodynamic turbulence~\cite{Kolmogorov1941}.
Second, systems with non-normal modes (non-orthogonal eigenvectors) display properties that are unexpected from linear calculations~\cite{camargo1998}. 
In fact, systems with non-normal modes even make it difficult to predict dynamics
when stable eigenmode branches are included in analyses~\cite{kim2010}. Third, linear stability analysis can miss crucial nonlinear instability effects, which come in several varieties. 

The most obvious variety of a nonlinear instability effect is that of subcritical turbulence in which no linear instabilities exist but turbulence is self-sustained 
given finite-amplitude seed perturbations. Subcritical turbulence is
common in hydrodynamics~\cite{manneville2008}.  While not as well-known in plasma physics,
several cases of subcritical plasma instabilities have been shown in the
literature~\cite{scott1990,scott1992,drake1995,nordman1993,waltz1985,itoh1996,highcock2012}.
The second variety of nonlinear instability includes cases in which a particular linear instability is present in a system, but the turbulence is maintained
by a nonlinear instability mechanism with different physical origin
than the linear instability mechanism. This has been explored in tokamak edge simulations in which
linear ballooning instability drive is overtaken in the saturated phase by a nonlinear drift-wave drive~\cite{zeiler1996,zeiler1997,scott2002,scott2003,scott2005}.
Finally, it is often found that a particular linear instability is enhanced, depressed, and/or  modified in the saturated phase by a nonlinear instability with a similar mechanism 
as the linear instability. In some of these cases nonlinear wavenumber transfers can increase or cause drive~\cite{biskamp1995,korsholm1999}, while in other cases zonal flow effects
decrease drive~\cite{dimits2000,ernst2004}. 

In order to avoid the pitfalls of relying too heavily on linear stability calculations in forming conclusions on turbulence characteristics, it is useful to perform turbulent simulations
and diagnose them with energy dynamics analyses. Energy dynamics analyses track energy input into turbulent fluctuations and energy dissipation out of them. 
They also track conservative energy transfer
between different energy types (e.g. from potential to kinetic energy) and between different scales, waves, or eigenmodes of a system. 
In all, energy dynamics analysis can be used as a post-processing
tool to characterize simulation turbulence in order to gain insight into underlying physical processes. 

In this study, a simulation of a
two-fluid Braginskii model of turbulence in the Large Plasma Device (LAPD) is subjected to such an energy dynamics analysis. This reveals that a nonlinear instability drives and maintains
the turbulence in the steady state saturated phase of the simulation. While a linear resistive drift wave instability resides in the system, the nonlinear drift wave instability dominates
when the mean fluctuation amplitude is over a few percent of the equilibrium value. The primary linear instability is the resistive drift wave which has a positive linear
growth rate for low but finite $k_\parallel$.  However, the saturated state of the simulated turbulence is strongly dominated by flute-like ($k_\parallel = 0$) fluctuations in density and potential.
The flute-like fluctuation spectrum is generated by a nonlinear instability.  The nonlinear instability is identified by its energy growth rate spectrum, which
varies significantly from the linear growth rate spectrum.  If $k_\parallel=0$ fluctuations are removed from the simulation (while retaining zonal flows), 
the saturated turbulent state is qualitatively and quantitatively different and much less consistent with experimental measurement.

\section{The drift wave model}
\label{dw_model}

A Braginskii-based fluid model~\cite{Braginskii1965} is used to simulate drift wave turbulence in LAPD using the BOUT++ code~\cite{dudson2009}. 
The evolved variables in the model are the plasma density, $N$, the electron fluid parallel velocity $\vpe$, the potential vorticity $\varpi \equiv \gradperp \cdot (N_0 \gradperp \phi)$,
and the electron temperature $T_e$. The ions are assumed cold in the
model ($T_i = 0$), which eliminates ion temperature gradient drive,
and sound wave effects are neglected. Details of the simulation code and derivations of the model
may be found in previous LAPD verification and validation studies~\cite{Popovich2010a,Popovich2010b,Umansky2011,friedman2012}, 
although electron temperature fluctuations were not included in those studies.

The equations are developed with Bohm normalizations: lengths are
normalized to the ion sound gyroradius $\rho_s$, times to the ion
cyclotron time $\omega_{ci}^{-1}$, velocities to the sound speed $c_s$, densities to the equilibrium peak density, and electron
temperatures and potentials to the equilibrium peak electron temperature. These normalizations are constants (not functions of radius) and are calculated from these reference values:
the magnetic field is $1$ kG, the ion unit mass is $4$, the peak density is $2.86 \times 10^{12}$ cm$^{-3}$, and the peak electron temperature
is $6$ eV. The equations are:

\beqar
\label{ni_eq}
\pdt N = - {\mathbf v_E} \cdot \grad N_0 - N_0 \gradpar \vpe + \mu_N \gradperp^2 N + S_N + \{\phi,N\}, \\
\label{ve_eq}
\pdt \vpe = - \fmie \frac{T_{e0}}{N_0} \gradpar N - 1.71 \fmie \gradpar T_e  + \fmie \gradpar \phi - \nue \vpe + \{\phi,\vpe \}, \\
\label{rho_eq}
\pdt \varpi = - N_0 \gradpar \vpe  - \nuin \varpi + \mu_\phi \gradperp^2 \varpi + \{\phi,\varpi \}, \\
\label{te_eq}
\pdt T_e = - {\mathbf v_E} \cdot \grad T_{e0} - 1.71 \frac{2}{3} T_{e0} \gradpar \vpe + \frac{2}{3 N_0} \kpe \gradpar^2 T_e  \nonumber \\
- \frac{2 m_e}{m_i} \nue T_e  + \mu_T \gradperp^2 T_e +  S_T + \{\phi,T_e\}.
\eeqar

\begin{figure}[!htbp]
\includegraphics[]{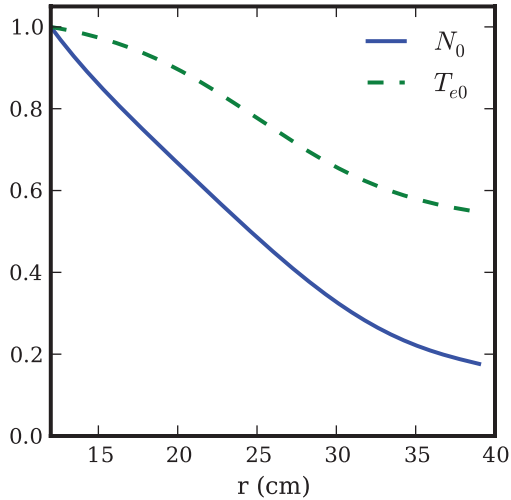}
\hfil
\caption{The profiles of density $N_0$ and electron temperature $T_{e0}$ used in the simulations normalized to their peak values of $2.86 \times 10^{12}$ cm$^{-3}$ and 
$6$ eV, respectively.}
\label{eq_profiles}
\end{figure}

In these equations, $\mu_N$, $\mu_T$, and $\mu_\phi$ are artificial diffusion and viscosity coefficients used for subgrid dissipation. They are large enough to allow saturation
and grid convergence~\cite{friedman2012}, but small enough to allow for turbulence to develop. In the simulations, they are all given the same value of $1.25 \times 10^{-3}$ in Bohm-normalized units. 
This is the only free parameter in the simulations. All other parameters such as the electron collisionality $\nue$, ion-neutral
collisionality $\nuin$, parallel electron thermal conductivity $\kpe$, and mass ratio $\fmie$ are calculated from the experimental parameters.
There are two sources of free energy: the density gradient due to the equilibrium density profile $N_0$, and the equilibrium electron temperature gradient in $T_{e0}$, both of which are
taken from experimental fits. $N_0$ and $T_{e0}$ are functions of only the radial cylindrical coordinate $r$, and they are shown in Fig.~\ref{eq_profiles}. 

The terms in Poisson brackets are the $E \times B$ advective nonlinearities, which are the only nonlinearities used in the simulations.
The numerical simulations are fully spatial in all three dimensions (as opposed to spectral) and use cylindrical annular geometry ($12<r<40$ cm). The radial extent used in the simulation
encompasses the region where fluctuations are above a few percent in the experiment.
The simulations use periodic boundary conditions in the axial ($z$) direction and Dirichlet boundary
conditions in the radial ($r$) direction for the fluctuating quantities. 

Simulations also use density and temperature sources ($S_n$ and $S_T$) in order to keep the equilibrium profiles from relaxing away from their experimental shapes. 
These sources suppress the azimuthal averages ($m=0$ component of the density and temperature fluctuations) at each time step. 
The azimuthal average of the potential $\phi$ is allowed to evolve in
the simulation, allowing zonal flows to arise.
The parallel current, which is often found explicitly in these equations is replaced here by $J_\para = - N_0 \vpe$. \\

\begin{figure}[!htbp]
\includegraphics[]{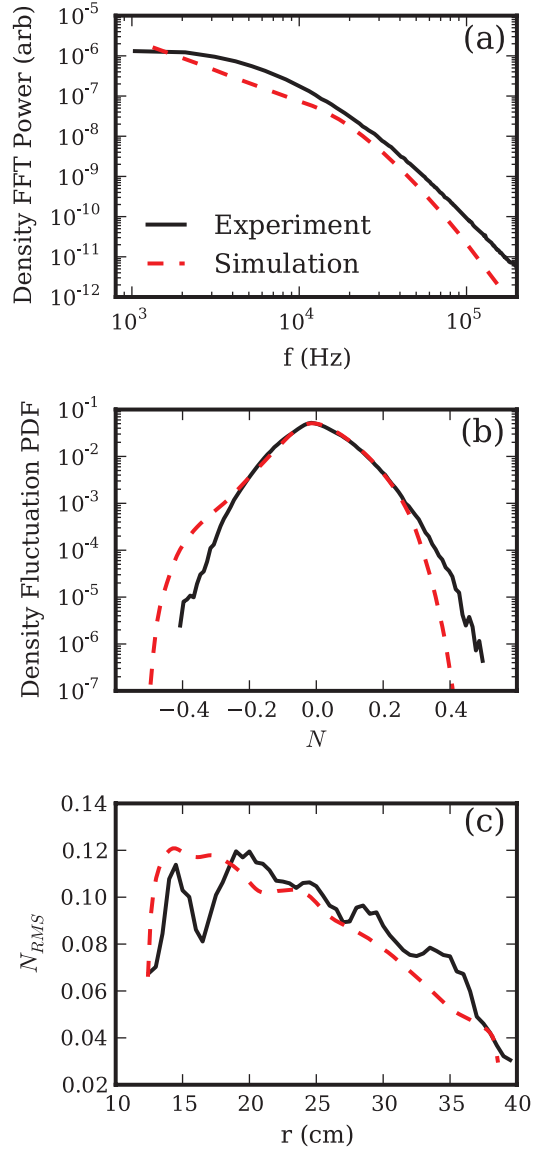}
\hfil
\caption{\textbf{a)} The power spectral density of the density fluctuations, showing the results from simulation versus experiment, \textbf{b)} the probability distribution function
of the density fluctuations, and \textbf{c)} the RMS amplitude of the density fluctuations as a function of radius.}
\label{validation}
\end{figure}

Some basic statistical properties of the density fluctuations of the
simulation are shown in Fig.~\ref{validation} and are  compared to the corresponding results from the experiment on which
this simulation is based. The simulation reproduces these characteristics of experimental measurements with rather good qualitative and quantitative accuracy.

\section{Energetics Machinery}
\label{sec_energetics_machinery}

In order to perform an energy dynamics analysis on the simulation, expressions for the energy and energy evolution must be derived from Eqs. \ref{ni_eq}-\ref{te_eq}.
To start, an expression for the normalized energy of the wave fluctuations in the drift wave model is defined as:

\beq
\label{energy_eq}
E = \frac{1}{2} \int_V  (N^2 + \frac{3}{2} T_e^2 + \frac{m_e}{m_i} \vpe^2 + N_0 (\gradperp \phi)^2 ) dV.
\eeq

The $N^2$ contribution is the potential energy due to density fluctuations, $\frac{3}{2} T_e^2$ is the electron temperature fluctuation potential energy,
$\frac{m_e}{m_i} \vpe^2$ is the parallel electron kinetic energy, and $N_0 (\gradperp \phi)^2$ is the $E \times B$ perpendicular kinetic energy.
These energy-like expressions are defined in this way so that they are conserved individually by their respective advective nonlinearities. While the physical energy contains
extra factors of $N_0$ and $T_{e0}$, the physical energy does not preserve the property of conservative nonlinearities of Eqs.~\ref{ni_eq}-\ref{te_eq} and therefore produces
a more complicated analysis. Analyzing an energy-like expression such as that in Eq.~\ref{energy_eq}, however, can be just as illuminating, and in this case simpler,
than analyzing the physical energy.
The expression in Eq.~\ref{energy_eq} will henceforth be referred to simply as the energy.

Furthermore, it is often most instructive to analyze the spectrally decomposed energy dynamics. 
To do this, each fluid field $(N,T_e,\vpe,\phi)$ at a given time is Fourier decomposed as 
$F(r,\theta,z) = \sum_{\vec{k}} f_{\vec{k}}(r) e^{i (m \theta + k_z z )}$,
where the subscript $\vec{k}$ represents the spectral wavenumbers, $(m,n)$. 
$m$ is the azimuthal wavenumber while $n$ is the axial integer wavenumber such that $k_z \equiv k_\para = 2 \pi n/l_z$. 
Note that the radial direction is not spectrally decomposed because the radial dependence of the profiles and differential operators complicates the analysis.
With this, the energy of each Fourier $\vec{k} = (m,n)$ mode is

\beq
\label{E_k}
E_{tot}(\vec{k}) = \frac{1}{2} \left< |n_{\vec{k}}|^2 + \frac{3}{2} |t_{\vec{k}}|^2 + \frac{m_e}{m_i} |v_{\vec{k}}|^2 + N_0 \left| \pdiff{\phi_{\vec{k}}}{r} \right|^2 + N_0 \frac{m^2}{r^2} |\phi_{\vec{k}}|^2 \right>,
\eeq

where the brackets $\left< \right>$ represent the radial integral: $\int_{r_a}^{r_b} r dr$. 
The energy evolution for each Fourier mode of each field has the form:

\beq
\label{dEdt_j}
\pdiff{E_{j}(\vec{k})}{t} = Q_{j}(\vec{k}) + C_{j}(\vec{k}) + D_j(\vec{k}) + \sum_{\vec{k}'} T_{j}(\vec{k},\vec{k}').
\eeq

The index $j$ stands for each field, ($n,t,v,\phi$), and the sum over $j$ gives the total energy evolution. 
Note that with the conventions used, the symbol $n$ denotes both the axial mode number as
well as the Fourier coefficient of the density fluctuation. The differences should be clear in context. The derivation of Eq.~\ref{dEdt_j} 
is given in the Appendix along with the full expressions for each of the parts. $T_{j}(\vec{k},\vec{k}')$ is the nonlinear energy transfer function that comes from the advective
nonlinearities.  It describes the nonlinear energy transfer rate of modes $\vec{k}'=(m',n')$ and $\vec{k}-\vec{k}'=(m-m',n-n')$ to the mode $\vec{k}=(m,n)$. 
In other words, a positive value of $T_{j}(\vec{k},\vec{k}')$ indicates that fluctuations
at wavenumber $\vec{k}$ gain energy from gradient fluctuations at wavenumber $\vec{k}'$ and flow fluctuations at wavenumber $\vec{k}-\vec{k}'$.
When summed over $\vec{k}'$ as in Eq.~\ref{dEdt_j}, the result is the total
nonlinear energy transfer into mode $\vec{k}$. 
Note that $\sum_{\vec{k},\vec{k}'} T_{j}(\vec{k},\vec{k}') = 0$ because the nonlinearities conserve energy individually in each of Eqs.~\ref{ni_eq}-\ref{te_eq}.
This is easily proven by the following identity:

\beq
\label{poisson_rln}
\int_\Omega q \{p,q\} \ d\Omega = \int_\Omega p \{p,q\} \ d\Omega = 0,
\eeq

which holds when boundary conditions are periodic or zero value as they are in the simulation. 
The fact that the advective nonlinearities conserve energy means that they can transfer energy between different Fourier modes
but they cannot change the energy of the volume-averaged fluctuations
as a whole. Only the linear terms can change the total energy of the fluctuations.
Other possible nonlinearities that do not conserve energy are not
included in the model equation set or in the simulations for simplicity of the energy analysis. 
Furthermore, it is convenient for the simulations to employ an energy conserving finite difference
scheme for the advective nonlinearities to reflect this analytic property of the equations. 
However, most common numerical advection schemes do not conserve energy for finite grid spacing. Therefore, an Arakawa advection scheme~\cite{arakawa1966} that conserves 
energy of the advected quantity is used for the nonlinear advection terms in the simulations.

The linear terms in Eqs. \ref{ni_eq}-\ref{te_eq} do not conserve energy individually or as a whole. The linear terms are broken up into three contributions in Eq.~\ref{dEdt_j}.
$D_{j}(\vec{k})$ represents nonconservative energy dissipation due to collisions, artificial diffusion and viscosity, and the density and temperature sources.
Each contribution to $D_j(\vec{k})$ is negative, and the exact expressions are given in the Appendix. 
$C_j(\vec{k})$ contains the linear terms dubbed ``transfer channels''~\cite{scott2002}. They are rewritten here:

\beqar
C_n(\vec{k}) & = & Re \left\{ \left< - i k_z N_0 v_{\vec{k}} n_{\vec{k}}^* \right> \right\}
\label{Cnk} \\
C_v(\vec{k}) & = & Re \left\{ \left< - i k_z N_0 n_{\vec{k}} v_{\vec{k}}^* + i k_z N_0 \phi_{\vec{k}} v_{\vec{k}}^*  - 1.71 i k_z T_{e0} t_{\vec{k}} v_{\vec{k}}^*  \right> \right\}
\label{Cvk} \\
C_\phi(\vec{k}) & = & Re \left\{ \left< i k_z N_0 v_{\vec{k}} \phi_{\vec{k}}^* \right> \right\}
\label{Cpk} \\
C_t(\vec{k}) & = & Re \left\{ \left< - 1.71 i k_z T_{e0} v_{\vec{k}} t_{\vec{k}}^* \right> \right\}
\label{Ctk}
\eeqar

Notice that $C_n(\vec{k}) + C_v(\vec{k}) + C_\phi(\vec{k}) + C_t(\vec{k}) = 0$, which is most clearly seen upon conjugation of $C_v(\vec{k})$ inside the real part operator.
This is the reason why these terms are called transfer channels. They represent the transfer
between the different types of energy of the different fields ($N,\phi,T_e \leftrightarrow v_{\para e}$), but taken together, they do not create or dissipate total
energy from the system. The only energy field transfer in this system occurs through the parallel electron velocity (parallel current) dynamics. There is no direct transfer between
the state variables $N, \phi,$ and $T_e$.  Altogether, the coupling through the parallel current is called the
adiabatic response. It is an essential part of both the linear and nonlinear
drift wave mechanisms~\cite{scott2002,scott2005}. The adiabatic response moves energy from the pressure fluctuations to the perpendicular flow through the parallel current. \\

Finally, the $Q_j(\vec{k})$ terms represent the nonconservative energy sources. They are rewritten here:

\beqar
Q_n(\vec{k}) & = & Re \left\{ \left< -\frac{i m}{r} \pdr N_0 \phi_{\vec{k}} n_{\vec{k}}^*  \right> \right\}
\label{Qnk} \\
Q_v(\vec{k}) & = & Re \left\{ \left<  i k_z \frac{N_0^2 - T_{e0}}{N_0} n_{\vec{k}} v_{\vec{k}}^* + i k_z (1 - N_0) \phi_{\vec{k}} v_{\vec{k}}^* + 1.71 i k_z (T_{e0} -1) t_{\vec{k}} v_{\vec{k}}^*  \right> \right\}
\label{Qvk} \\
Q_\phi(\vec{k}) & = & 0
\label{Qpk} \\
Q_t(\vec{k}) & = & Re \left\{ \left< -\frac{3}{2} \frac{i m}{r} \pdr T_{e0} \phi_{\vec{k}} t_{\vec{k}}^*  \right> \right\}
\label{Qtk}
\eeqar

$Q_n(\vec{k})$ is the energy extraction from the equilibrium density profile into the density fluctuations. 
This term may have either sign depending on the phase relation between $\phi_{\vec{k}}$ and $n_{\vec{k}}$, 
so it can in fact dissipate fluctuation potential energy from the system as well as create it
at each $\vec{k}$. $Q_t(\vec{k})$ is completely analogous to $Q_n(\vec{k})$ but for the temperature rather than the density. 
$Q_v(\vec{k})$ is parallel kinetic energy extraction or dissipation.
The sources of these terms are the equilibrium gradients, which is evident because if the profiles
were flat ($N_0=T_{e0}=1$), all $Q(\vec{k})$ would vanish.
Moreover, the particular normalization of Eqs.~\ref{ni_eq}-\ref{te_eq} combined with the choice of energy definition (Eq.~\ref{energy_eq}) causes the non-zero $Q_v(\vec{k})$.
The energy dynamics of the physical energy does not have a finite $Q_v(\vec{k})$. However, with the energy-like expression analyzed here $Q_v(\vec{k})$ is small compared to the 
parallel velocity dissipation $D_v(\vec{k})$, meaning that there is no net energy entering the system directly through the parallel velocity as expected. 
The energy expression used here is therefore a good proxy for the physical energy with the added benefit that the expression used here conserves the nonlinearities.

\section{Nonlinear Energy Dynamics}
\label{nl_en_dynamics_sec}

\subsection{Energy Spectra}

\begin{figure}[!htbp]
\includegraphics[]{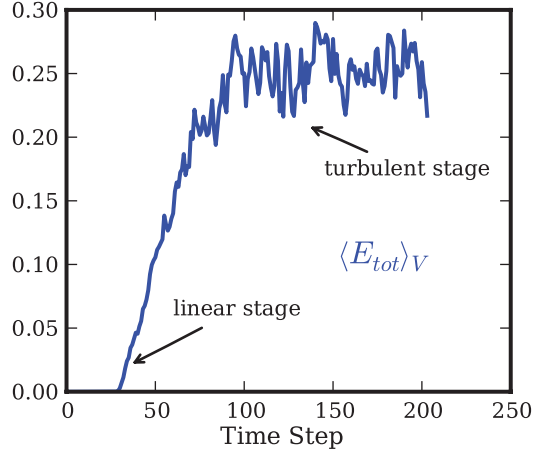}
\hfil
\caption{Time evolution of the volume-averaged total energy. Each time step is $400/\omega_{ci} \sim 170 \mu s$}
\label{time_evolution}
\end{figure}

\begin{figure}[!htbp]
\includegraphics[]{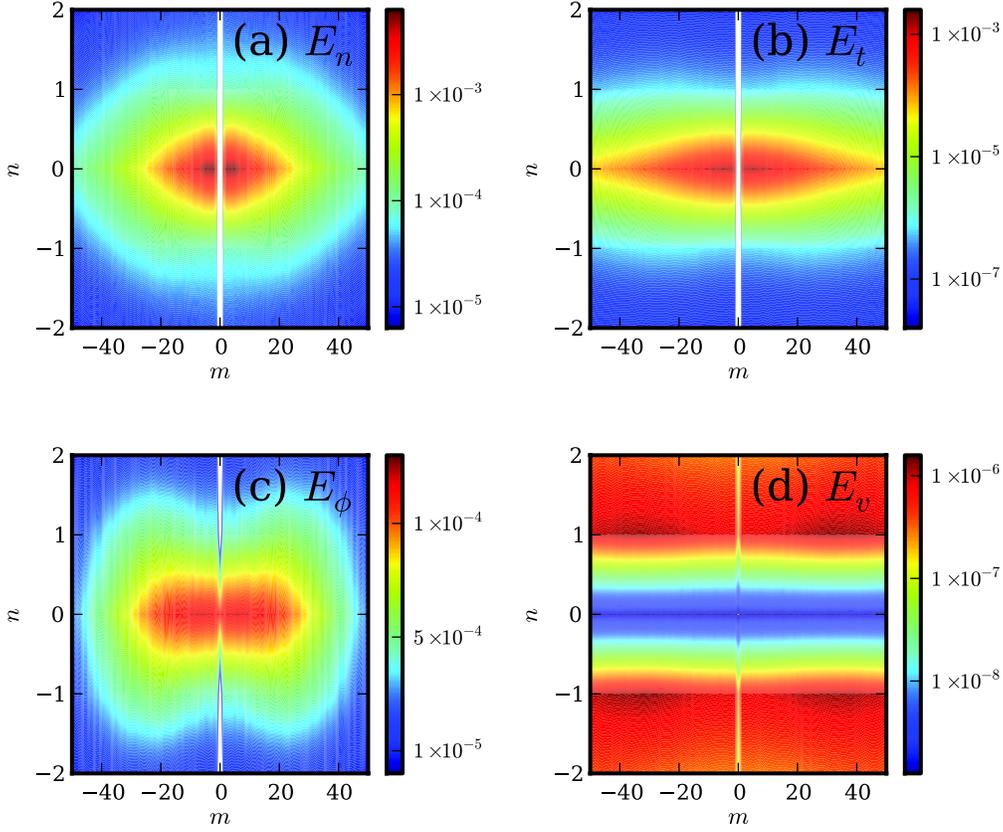}
\hfil
\caption{\textbf{a)} $E_n(\vec{k})$, \textbf{b)} $E_t(\vec{k})$, \textbf{c)} $E_\phi(\vec{k})$, and \textbf{d)} $E_v(\vec{k})$ in the $m-n$ plane averaged over time during the saturated turbulent phase.
Note the different scales used on each figure and also that the vertical white lines at $m=0$ in \textbf{a)} and \textbf{b)} are due to the density and temperature sources which subtract out this
component of the fluctuations. The zonal flow, defined as the $n=0, m=0$ component of the $E \times B$ velocity, has an energy of magnitude of $1.4 \times 10^{-4}$ as seen in \textbf{c)}.}
\label{energy_figures}
\end{figure}

Figure~\ref{time_evolution} shows the time evolution of the total energy of the fluctuations. The simulation starts with a random initial perturbation, and the fluctuations grow
exponentially due to the linear drift wave instability until the energy level reaches about $0.01$, where the energy is fairly equally divided between $n=0$ and $n= \pm 1$ modes. 
Then, the nonlinear instability takes over and the fluctuation energy continues to grow
until reaching saturation. All analysis shown below is done by time averaging over the saturated (turbulent) stage.
The turbulent spectral energy, defined in Eq.~\ref{E_k} is shown in Fig.~\ref{energy_figures}. The energy is broken up into its different types (e.g. perpendicular kinetic energy: $E_\phi$).
There are a few clear nonlinear properties
seen in these figures. The first is that the energy is located in different spectral regions for the different energy types. This has to be a nonlinear effect because the linear eigenmodes
are Fourier modes in the azimuthal and axial directions and all fields grow at the same rate for an eigenmode. 
Another property unexpected from linear stability analysis is that most of the potential and perpendicular kinetic energy ($E_n$, $E_t$, and $E_\phi$) is contained in $n=0$ ($k_\para = 0$) 
structures, which are often called flute modes. Previous studies pointed out this flute mode dominance in LAPD simulations~\cite{rogers2010,Umansky2011}. The study by Rogers et al.~\cite{rogers2010},
however, used a momentum source that produced a large radial electric field, possibly leading to a dominating Kelvin-Helmoltz instability at $k_\para = 0$.
Such a feature is unexpected in this study because there is no $n=0$ linear instability present in the system, which is confirmed by eigenvalue calculations~\cite{Popovich2010a}.
The only linear instability of the system is the
linear resistive drift wave instability, which requires finite $n$ to provide the phase shift and state variable coupling to drive the waves unstable. Perhaps equally unexpected is the complete
lack of parallel kinetic energy in the $n=0$ range. The $E_v$ spectrum looks like a traditional linear drift wave spectrum, but does not match the other fields, which is atypical of
linear drift waves.


\subsection{Description and Evidence for the Nonlinear Instability}

\begin{figure}
\includegraphics[width=3in,height=2.5in]{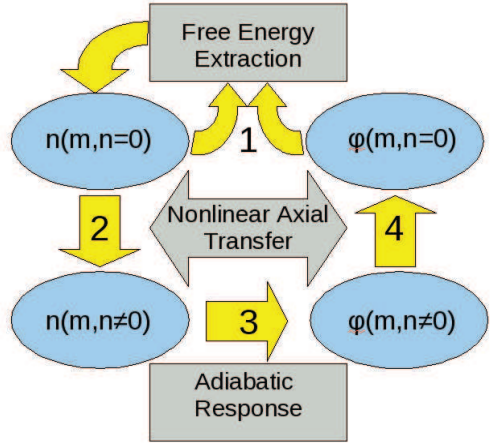}
\hfil
\caption{Diagram of the nonlinear instability process that drives flute modes.}
\label{nl_instability_diagram}
\end{figure}

The flute mode dominance has to be a nonlinear effect because linear drift waves require finite $n$. However, the cause of the flute dominance is not simple cascade dynamics, a secondary 
instability, nor a flow-driven flute-like instability such as
Kelvin-Helmholtz or interchange. Rather, the cause is a primary
nonlinear instability as has been reported in previous simulations of
plasma edge turbulence~\cite{biskamp1995,drake1995}. 
This nonlinear instability is a multi-step process that is outlined in Fig.~\ref{nl_instability_diagram}.
In the first step, $n=0$ density and potential fluctuations \emph{nonconservatively} draw energy from the equilibrium density gradient as prescribed 
by $Q_n(m,n=0)$ defined in Eq.~\ref{Qnk}, and feed
this energy into the $n=0$ density fluctuations only. The nonconservative linear terms, after all, can only inject, dissipate, or transfer energy at one wavenumber at a time, so it takes
$n=0$ fluctuations to nonconservatively inject energy into $n=0$
fluctuations. Note also that the temperature fluctuations work in the same way as the density fluctuations, and one could replace $n(m,n)$ (the spectral density component)
with $t(m,n)$ (the spectral temperature component) in the diagram. The density and temperature are analogous and work in parallel, however the temperature fluctuations are a few times smaller
than the density fluctuations and provide weaker drive because
parallel heat conduction strongly dissipates the temperature
fluctuations. This is why the diagram highlights the density
contribution (this observation is consistent with other work in edge turbulence~\cite{zeiler1997}).
 This nonconservative injection does not occur for
infinitesimal perturbations; a finite-amplitude $n=0$ seed
perturbation is required.  In the simulations, this seed is provided
by nonlinear transfer from $n=1$ drift wave fluctuations which
dominate the linear phase of the turbulence simulation.

In the second step of the diagram, these $n=0$ density fluctuations
\emph{conservatively} transfer energy to $n \ne 0$ density fluctuations by the nonlinear $T_n(\vec{k},\vec{k}')$ transfer process. 
The third step involves the transfer at finite $n$ from the density
fluctuations to the potential fluctuations by way of the parallel current in the adiabatic response. The $C_j(\vec{k})$ terms describe this adiabatic response.
Fourth and finally, the $T_\phi(\vec{k},\vec{k}')$ interaction conservatively transfers energy from
$n \ne 0$ to $n=0$ potential $\phi$ fluctuations in inverse fashion, providing the necessary potential flute structures for the first step.

\begin{figure}
\includegraphics[]{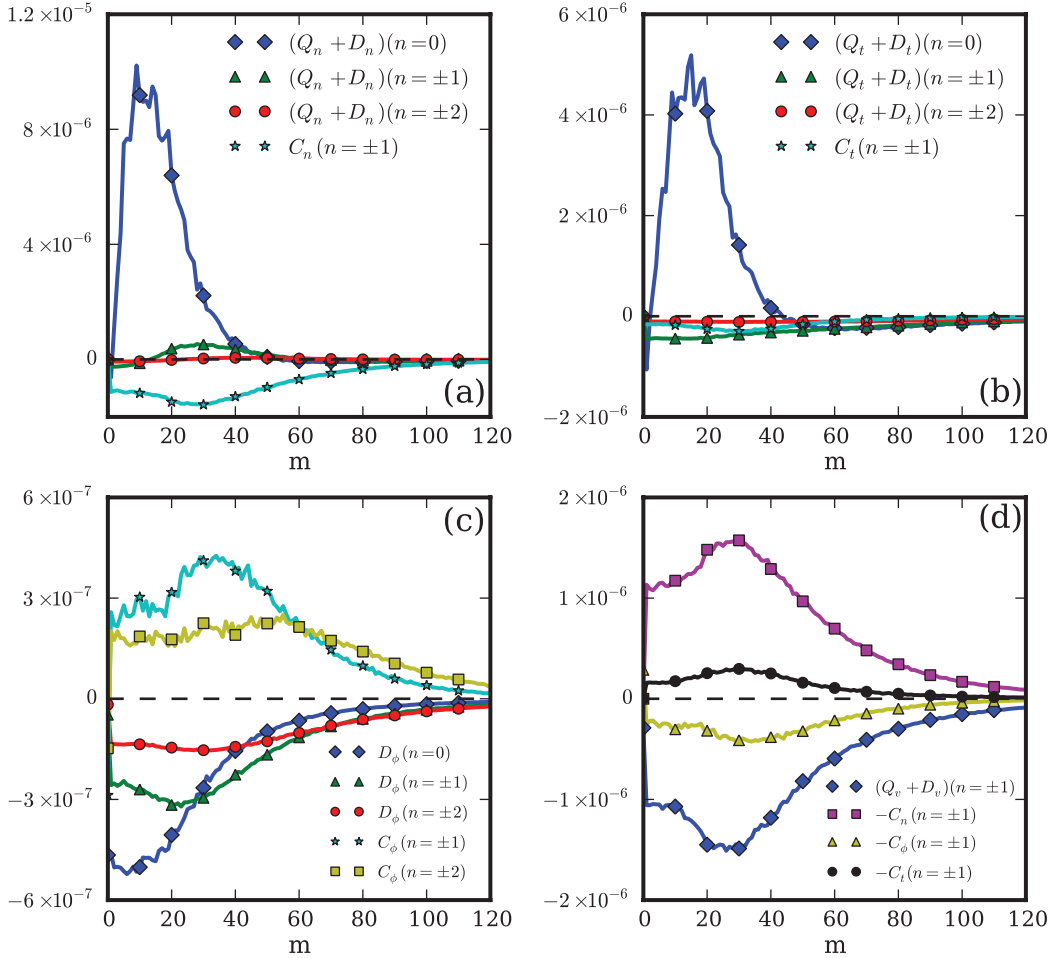}
\hfil
\caption{\textbf{a)} The solid curves quantify the energy dynamics of the density potential energy averaged over time during the saturated turbulent phase. The notation $n \pm 1$ represents the
summation over the $n=1$ and $n=-1$ curves. \textbf{b)} The energy dynamics of the temperature potential energy, \textbf{c)} the perpendicular kinetic energy, and \textbf{d)} the parallel
dynamics (adiabatic response). The contributions to $C_v(\vec{k})$ in \textbf{d)} are broken up with $C_v(\vec{k}) = -C_n(\vec{k}) - C_t(\vec{k}) - C_\phi(\vec{k})$. 
The density and temperature sources are neglected in \textbf{a)} and \textbf{b)} respectively. They only contribute at $m=0$.}
\label{nc_dynamics_figures}
\end{figure}

The evidence for the dominance of this mechanism is best shown with
help from the energy dynamics machinery derived in section~\ref{sec_energetics_machinery}. Figure~\ref{nc_dynamics_figures} summarizes the effects of the nonconservative linear terms, which are
fully responsible for injecting energy into the fluctuations. Figure~\ref{nc_dynamics_figures}(a) shows the $E_n$ dynamics separated into different parallel wavenumbers and plotted
against the azimuthal wavenumber $m$. Clearly, most of the energy is injected into $n=0$ density structures, while only a small amount of energy is injected into
$n= \pm 1$ structures despite the fact that the linear instability is active only at $n \ne 0$. The large positive $Q_n + D_n$ (injection plus dissipation) at $n=0$ provides evidence for the first
step of the diagram in Fig.~\ref{nl_instability_diagram}. Note however that the dissipation from the source, which acts entirely at $m=0$, is neglected in this figure because it is so large 
(about $5 \times 10^{-5}$) that it would compress the other lines too much.
Modes with $|n| \ge 2$, on the other hand, play a negligible role in density injection, dissipation, and transfer. 
Furthermore, all of the net energy injected into the density fluctuations ($Q_n + D_n$) is transferred out ($C_n$) to
the parallel current (electron velocity), which only occurs at finite $n$, almost entirely at $n = \pm 1$. The net change of $E_n$, which is the sum $Q_n + D_n + C_n$ over all $m$ and $n$
is approximately zero because this analysis is averaged over the steady state turbulence, although this is not so evident in Fig.~\ref{nc_dynamics_figures}a without the source dissipation. 
The necessary balance implies, as will be proven later, that the nonlinearities transfer energy from $n=0$ to $n = \pm 1$ modes, where that energy can then
be transferred to the parallel current.

Figure~\ref{nc_dynamics_figures}(b) shows the temperature potential energy dynamics. Again flute structures inject energy into the fluctuations, but unlike in the density case, $n = \pm 1$ modes
dissipate more energy than they inject. Moreover, the small value of $C_t$ reveals that the temperature fluctuations inject only a small amount of energy into the parallel current compared
to the density fluctuations. Despite the fact that the equilibrium temperature gradient is nearly as steep as the density gradient at its steepest point, 
its free energy is not used efficiently by the waves
in the sense that it is largely dissipated before being transferred to the electrostatic potential. The reason for the difference
between the density and temperature responses is the extra dissipation routes for the temperature fluctuations, namely, the parallel heat conduction and electron-ion heat exchange.
One should therefore be careful in assuming that adding free energy sources to an analysis will automatically increase instability drive.
The same type of result was seen in a study of tokamak edge turbulence~\cite{zeiler1997}, although there, the temperature fluctuations were even more dissipative than in this study in that they
actually drew energy from the parallel current.

Next, Fig.~\ref{nc_dynamics_figures}(c) illustrates the perpendicular kinetic energy dynamics provided by the electrostatic potential $\phi$. 
Since there is no free energy source for the potential ($Q_\phi=0$),
the potential fluctuations derive their energy from the parallel current through the $C_\phi$ transfer channel, which is positive everywhere and only finite for finite $n$. Otherwise, ion-neutral
collisions and viscosity dissipate energy from the potential fluctuations as shown by the $D_\phi$ curves. An interesting detail seen in this figure is that modes with $|n| > 1$ actually
contribute to the transfer channel and dissipation, whereas these modes are negligible for the other fields. 

The last piece, the parallel dynamics, also called the adiabatic response, is displayed in Fig.~\ref{nc_dynamics_figures}(d). 
The primary effect of the adiabatic response is to take energy from the density
fluctuations and transfer it to the potential fluctuations, which only occurs at finite parallel wavenumber. This effect corresponds to the third step in Fig.~\ref{nl_instability_diagram}.
Moreover, resistivity dissipates a substantial portion of this
parallel kinetic energy and, although not evident from this figure, provides
the primary phase shift mechanism between the density and potential that allows for instability. The temperature fluctuations also provide energy to the potential fluctuations
through this response ($-C_t$), although it is much weaker than the density fluctuation route.

\begin{figure}[!htbp]
\includegraphics[]{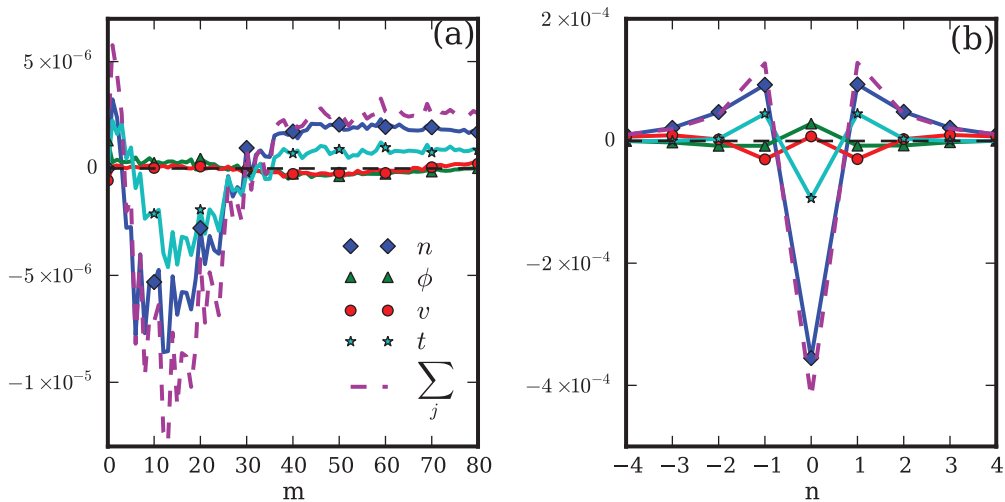}
\hfil
\caption{\textbf{a)} Conservative nonlinear energy transfer functions $T_j(\vec{k},\vec{k}')$ summed over $\vec{k}'$ and $n$. 
The line labeled $\phi$ represents the function $\sum_{\vec{k}',n} T_\phi(\vec{k},\vec{k}')$,
which is a function of $m$. The line labeled $\sum_j$ is the total energy transfer (the sum over the other lines).
\textbf{b)} Transfer functions summed over $\vec{k}'$ and $m$. Note that $T_v(\vec{k},\vec{k}')$ is multiplied by $20$ in both figures to make it visibly different from zero.}
\label{conservative_transfers}
\end{figure}

Steps 2 and 4 in Fig.~\ref{nl_instability_diagram} come not from the nonconservative linear terms in the equations, but from the conservative nonlinear advective terms.
The interactions described by the advective nonlinearities are in the nonlinear transfer functions: $T_{j}(\vec{k},\vec{k}')$ in Eq.~\ref{dEdt_j}.
It is difficult to study the $T_{j}(\vec{k},\vec{k}')$ functions because they are four dimensional functions of $(m,n,n',m')$, 
which makes visualization challenging. It is therefore convenient to sum over various transfers or look at specific wavenumber transfers of interest. The most easily decipherable
results that complement the results of Fig.~\ref{nc_dynamics_figures} are shown in Fig.~\ref{conservative_transfers}. First, Fig.~\ref{conservative_transfers}(a) sums the transfer functions over
$(n,m',n')$, leaving only a function of $m$, which illustrates the aggregate azimuthal mode
number transfers. Note that the sum of each individual curve over $m$ is zero because the nonlinearities are conservative. Positive values in Fig.~\ref{conservative_transfers}(a)
indicate energy transfer into structures with azimuthal mode number $m$, while negative values indicate energy transfer out of structures with corresponding mode number $m$. 
The density and temperature nonlinearities
are qualitatively similar in that they cause both forward and inverse transfer out of the wavenumbers that nonconservatively inject the most energy. The potential (polarization) and parallel velocity 
nonlinearities cause inverse transfer to low wavenumbers. 

Figure~\ref{conservative_transfers}(b) displays the conservative transfers summed over $(m,m',n')$, leaving only a function of $n$, which describes transfer into and out
of different parallel modes. This is the figure which provides evidence for steps 2 and 4 of the instability diagram.
Now, as expected from step 2 of the diagram and foreshadowed by Fig.~\ref{nc_dynamics_figures}(b), 
density potential energy is aggregately transferred from $n=0$ flute modes into $n \ne 0$ modes, specifically $n = \pm 1$. This can be called a direct
transfer in analogy with the terminology used for hydrodynamic wavenumber cascades. The temperature fluctuations have the same transfer trends 
as the density fluctuations, while the parallel velocity exhibits direct transfer, although from $|n|=1$ to higher modes since there is never any $n=0$ energy in the parallel velocity.
The potential fluctuations, on the other hand, exhibit inverse parallel wavenumber transfer (step 4 of the diagram),
populating $n=0$ potential structures. This nonlinear transfer is the only way to drive energy into
$n=0$ potential structures because $Q_\phi=0$. That completes the evidence for the nonlinear instability picture along with further details of both the conservative and nonconservative energy dynamics.

\subsection{The Global Energy Injection and Dissipation Picture}

\begin{figure}[!htbp]
\includegraphics[]{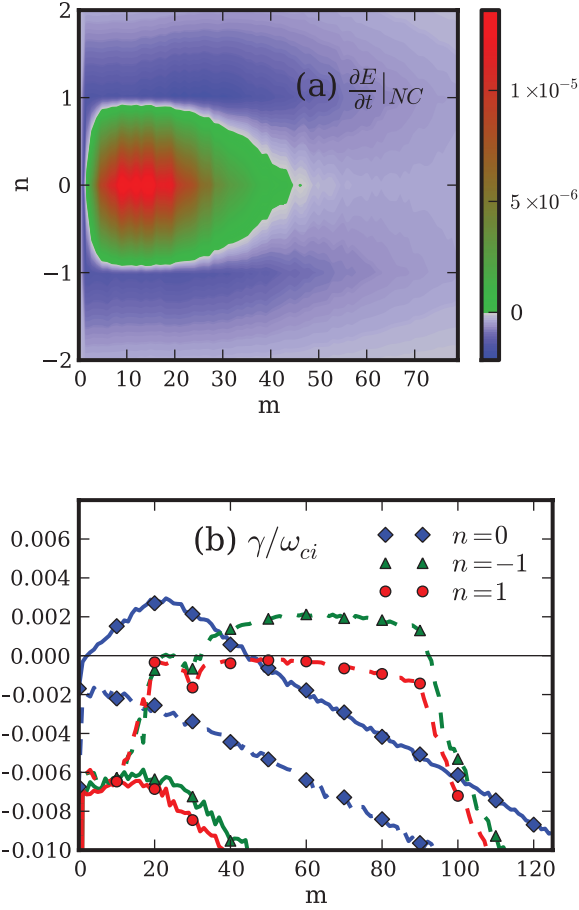}
\hfil
\caption{\textbf{a)} The total spectral nonconservative energy injection $\pdiff{E(\vec{k})}{t} \big|_{NC}$ and
\textbf{b)} the spectral nonconservative growth rate spectrum $\gamma_{T}(\vec{k})$ of the turbulence 
compared to the linear growth rate spectrum, $\gamma_L(\vec{k})$. 
The solid lines represent $\gamma_{T}(\vec{k})$ which is calculated using Eq.~\ref{gamma_def} averaged over the saturated turbulent phase, 
while the dashed lines represent $\gamma_L(\vec{k})$ and are calculated with the same equation using the linear phase of the simulation.}
\label{dEdt_tot_fig}
\end{figure}

The details of the energy dynamics given above are important but can obscure the most significant results. Specifically, Fig.~\ref{nc_dynamics_figures} contains a lot of details
that can be contracted by summing over the different energy types.
Figure~\ref{dEdt_tot_fig} does this, showing the total spectral nonconservative energy dynamics.
Figure~\ref{dEdt_tot_fig}(a), which is a plot of the nonconservative rate of change of the total energy, $ \pdiff{E(\vec{k})}{t} \big|_{NC} = \sum_j Q_j(\vec{k}) + C_j(\vec{k}) + D_j(\vec{k})$, 
reveals a global picture in wavenumber space of
where the total energy is injected into the system and where it is dissipated. Namely, energy injection occurs on average at $(n=0, 3<m<45)$, 
while it is dissipated everywhere else including at $n=\pm 1$ for all $m$. It is obvious that the nonlinear wavenumber transfers must take energy from the injection region to the dissipation
region on average, and that is consistent with what was shown in Fig.~\ref{conservative_transfers}.
This further reveals a picture quite different than what one would expect from the standard picture of plasma turbulence in which energy is 
injected where the \emph{linear} growth rate is positive and dissipated
where it is negative. The picture here is quite the opposite -- energy is injected where there is no linear instability and dissipated in part where there is one.
To clarify this point, the linear growth rate $\gamma_L(\vec{k})$ versus turbulent growth rate $\gamma_{T}(\vec{k})$ spectra are shown in Fig.~\ref{dEdt_tot_fig}(b). 
The growth rates are calculated using:

\beq
\label{gamma_def}
\gamma(\vec{k}) = \left( \sum_j Q_j(\vec{k}) + C_j(\vec{k}) + D_j(\vec{k}) \right)/\left( 2 \sum_j E_j(\vec{k}) \right).
\eeq

The turbulent growth rate spectrum, simply means that Eq.~\ref{gamma_def} is calculated using the terms from the saturated turbulent phase of the simulation.
Note that the linear growth rate is positive for $(n=-1, 35<m<95)$ and negative everywhere else. The turbulent growth rate is positive only
for $(n=0, 3<m<45)$. The linear and turbulent spectral injection regions do not even overlap. Seemingly, the linear physics is completely washed out in the turbulent state.

\section{Linear Versus Nonlinear Instability Drive}

\begin{figure}[!htbp]
\includegraphics[]{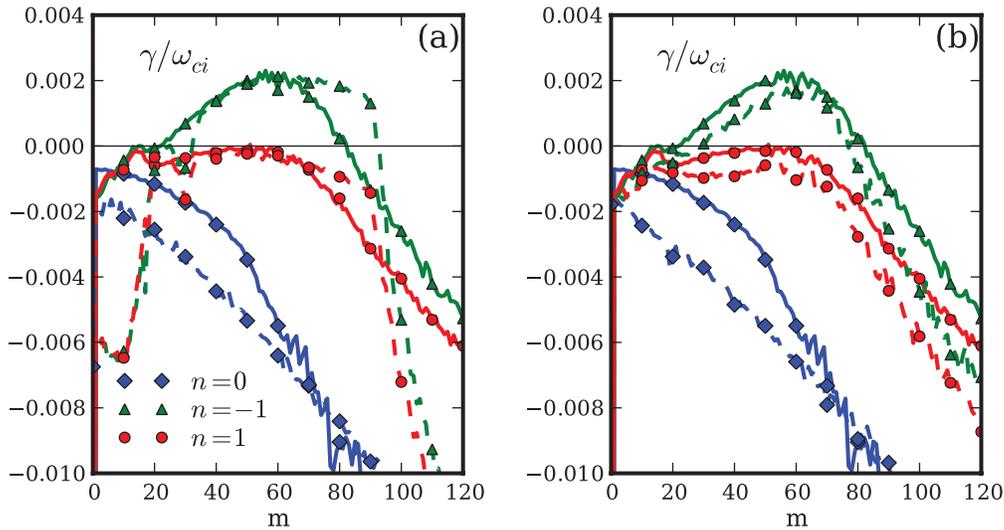}
\hfil
\caption{\textbf{a)} The turbulent growth rate spectrum $\gamma_T(\vec{k})$ with $n=0$ density and potential components removed from the 
simulation compared to the linear growth rate spectrum $\gamma_L(\vec{k})$.
\textbf{b)} The solid lines are the same $\gamma_T(\vec{k})$ spectrum as the solid lines in (a), but the dashed lines are the turbulent growth rate spectrum when the zonal flows are retained
in the simulation.}
\label{gamma_no_n0_figs}
\end{figure}

Although the nonlinear flute mode dynamics present a clear case of nonlinear instability, the $n \ne 0 \leftrightarrow n=0$ energy path is not a necessary feature of nonlinear drift wave
instabilities, which is clear in tokamak studies of drift wave turbulence~\cite{zeiler1996,zeiler1997,scott2002,scott2003,scott2005}. Furthermore, the periodic axial boundary conditions
used in the LAPD turbulence simulation are obviously unphysical, and more realistic boundary conditions may change the parallel dynamics disallowing an exact $n \ne 0 \leftrightarrow n=0$ path.

In essence, it is interesting to test the robustness of nonlinear instability in this system. In particular, how important are the idealized flute modes to the nonlinear instability?
They are after all, not essential to the otherwise similar nonlinear drift-like instabilities in the tokamak edge simulations~\cite{zeiler1996,zeiler1997,scott2002,scott2003,scott2005}.
Now, there are a few ways to eliminate the flute modes in the simulation such as
eliminating one of the nonlinearities that is essential to the nonlinear instability process described in Fig.~\ref{nl_instability_diagram}. However, simply subtracting out the $n=0$
components of the density and potential at each simulation time step retains more of the physics that may be essential to cause nonlinear instability. The energy dynamics of such a simulation,
which are succinctly summarized by the growth rate spectrum,
are shown in Fig.~\ref{gamma_no_n0_figs}. Interestingly, with no $n=0$
fluctuations, the turbulent growth rate spectrum $\gamma_T(\vec{k})$ is nearly identical to the linear growth rate spectrum $\gamma_L(\vec{k})$, as seen in
Fig.~\ref{gamma_no_n0_figs}(a). It is noted that subtracting out the $n=0$ potential component removes the zonal flow ($m=n=0$) from the system, providing a possible explanation for the large change
in behavior of the turbulent growth rate spectrum. However, this hypothesis is dispelled by the analysis of a simulation in which only the $n=0, m \ne 0$ potential components are removed while 
the zonal flow is left intact. The growth rate spectrum of this simulation, shown in Fig.~\ref{gamma_no_n0_figs}b, reveals that the zonal flow plays a minimal role in the nonlinear instability
dynamics. The zonal flow simply decreases the growth rates by a small amount, causing no change to the qualitative picture. 

\begin{figure}[!htbp]
\includegraphics[]{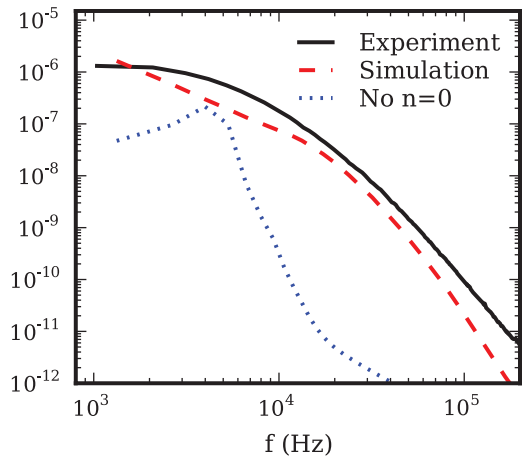}
\hfil
\caption{Comparison of the frequency spectra. Notice the spectra with $n=0$ components removed is not broadband, but has a clear peak, which is inconsistent with experiment.}
\label{freq_spectra3}
\end{figure}

The flute modes are therefore necessary for a nonlinear instability to overtake the linear instability in driving the turbulence, making this qualitatively different from the
tokamak edge studies. Furthermore, it is clear from this result that a 2D simplification
using a fixed parallel wavelength like the 2D Hasegawa-Wakatani model~\cite{hasegawa1983} does not support a nonlinear instability.
Nevertheless, one indication that the nonlinear flute-driven instability is important in reproducing experimentally consistent turbulence 
is that the turbulence of the simulation with the $n=0$ components removed becomes overly coherent. This can be seen in the
frequency spectrum, which is shown in
Fig.~\ref{freq_spectra3} compared with the experimental spectra and
the spectra of the standard nonlinear-instability-dominated
simulation. While the standard simulation with the $n=0$ modes retained compares
well with experimental data, the spectrum of the simulation with $n=0$ modes removed does not. It is not broadband, having a large peak, which is
inconsistent with experimental spectrum.  Apparently the $n=0$ modes and the nonlinear instability are important for reproducing experimentally relevant turbulence. 
A more direct test with realistic axial boundary conditions is left for a future study.

\section{Conclusion}

In contrast to experiments, simulations provide vast quantities of spatial information, and can therefore be useful in illuminating physical processes responsible for driving and saturating
turbulence. It is possible to get more than fluctuation levels, flux values, diffusivities, and spectra from simulations. The kind of energy analysis used in this study is one 
way in which detailed physics can be drawn from a turbulence simulation. Here, energy dynamics analysis shows a complex picture of turbulent energy injection, transfer, and dissipation. 
Such a picture was certainly not evident \emph{a priori}.
Other more advanced procedures such as eigenmode decompositions~\cite{baver2002} or proper orthogonal decompositions (POD)~\cite{hatch2011}, which are extensions of this procedure,
can reveal even more physical processes, especially those involving saturation. 

In this study, a partial spectral decomposition and energy dynamics analysis was sufficient to reveal the dominance of a nonlinear instability in driving and maintaining the turbulence.
The nonlinear instability works by driving $k_\parallel = 0$ pressure fluctuations using $k_\parallel = 0$ pressure and potential fluctuations to access the free energy pressure gradients.
These $k_\parallel = 0$ pressure fluctuations are \emph{not} driven by conservative nonlinear energy transfer from linear drift waves nor by some primary linear flow-driven instability.
The $k_\parallel = 0$ potential fluctuations are driven through the finite $k_\parallel$ adiabatic response in combination with forward and reverse axial wavenumber transfers. Not only
does the nonlinear instability require the $k_\parallel = 0 \leftrightarrow k_\parallel \ne 0$ transfer path to operate, but the simulation requires this to produce experimentally
consistent broadband turbulence. In the future, this study will be expanded to include different and more realistic axial boundary conditions -- including conducting plate boundaries --
to further test the importance of the flute modes in creating broadband turbulence. Furthermore, different operating conditions in LAPD, including those that produce large mean flows
will be simulated and studied to test for the emergence of new dominant instabilities, possibly nonlinear ones.
 
Understanding nonlinear instabilities is important because they can invalidate the use of quasilinear flux estimates and linear mixing length arguments of turbulent transport 
levels when linear instabilities are insignificant in the turbulent state. Simple rules for when nonlinear instabilities will act or overtake linear ones are needed, and one attempt at such
a rule has been made elsewhere for drift wave turbulence~\cite{scott2005}. That rule states that nonlinear instabilities will overtake linear instabilities when $\gamma_L < \omega_*$, which is true
for the parameters used in this study. However, more study of this rule and others is warranted, and will be important as long as linear calculations are used to inform predictions of turbulence.
Nevertheless, full nonlinear simulations and energy dynamics analyses
are most informative and should be used to obtain details of plasma
turbulence mechanisms.

\begin{acknowledgments}
This research was performed under appointment to the Fusion Energy Sciences Fellowship Program administered by Oak Ridge Institute for
Science and Education under a contract between the U.S. Department of Energy and the Oak Ridge Associated Universities. 
We would also like to thank Prof. Paul Terry and Dr. Ilon Joseph for many useful discussions on this topic.
\end{acknowledgments}

\appendix

\section{Explicit Calculation of the Energy Evolution of a Fourier Mode}

The energy evolution for each Fourier mode can be obtained by Fourier decomposing each of Eqs. \ref{ni_eq} - \ref{te_eq}
and then multiplying the density, electron parallel velocity, vorticity, and electron temperature equations by the complex conjugates of the 
density, velocity, potential, and temperature respectively, and integrating over space.
Adding the resulting equations gives the energy evolution of each Fourier mode.

Take the density equation as an example for this procedure. The decomposition for the density is:

\beq
\label{density_decomp}
N(r,\theta,z,t) = \sum_{\vec{k}} n_{\vec{k}}(r,t) e^{i(m \theta + k_z z)}.
\eeq

Recall that the subscript $\vec{k}$ is short for $(m,n)$ as the decomposition is a 2D Fourier decomposition in the azimuthal and axial directions, making the sum over $\vec{k}$ truly
a double sum over $m$ and $n$. Furthermore, positive and negative
$m$ and $n$ are included in the sums to ensure reality of $N$, which also requires that $n_{-\vec{k}} = n_{\vec{k}}^*$.
Similar decompositions are used for $\vpe$ and $\phi$. The density source is azimuthally symmetric, so it is decomposed as:

\beq
\label{source_decomp}
S_N(r,z,t) = \sum_{k_z} S_{N k_z}(r,t) e^{i k_z z}.
\eeq

Substituting this decomposition into Eq. \ref{ni_eq} gives:

\beqar
\label{density_eq_fourier}
\sum_{\vec{k}} \pdiff{n_{\vec{k}}}{t} e^{i(m \theta + k_z z)} = \sum_{k_z} S_{N k_z} e^{i k_z z} + \nonumber \\
\sum_{\vec{k}} \left[ -\frac{i m}{r} \pdr N_0 \phi_{\vec{k}} - i k_z N_0 v_{\vec{k}} + \mu_N(\pdrr n_{\vec{k}} + \frac{1}{r} \pdr n_{\vec{k}} - \frac{m^2}{r^2} n_{\vec{k}}) \right] e^{i(m \theta + k_z z)} \nonumber \\
+ \frac{1}{r} \sum_{\vec{k},\vec{k}'} (i m n_{\vec{k}} \pdr \phi_{\vec{k}'} - i m' \pdr n_{\vec{k}} \phi_{\vec{k}'}) e^{i (m + m') \theta + i (k_z + k'_z) z}.
\eeqar

Multiplying through by $n_{\vec{k}''}^* e^{- i m'' \theta - i k''_z z}$ and integrating over space (and permuting primes) gives:

\beqar
\label{density_evolution}
\left< \pdiff{n_{\vec{k}}}{t} n_{\vec{k}}^* \right> = \left< S_{N k_z} n_{m=0,k_z}^* \right>  \nonumber \\
\left< -\frac{i m}{r} \pdr N_0 \phi_{\vec{k}} n_{\vec{k}}^* - i k_z N_0 v_{\vec{k}} n_{\vec{k}}^* + \mu_N( \pdrr n_{\vec{k}} + \frac{1}{r} \pdr n_{\vec{k}} - \frac{m^2}{r^2} n_{\vec{k}}) n_{\vec{k}}^*  \right> \nonumber \\
+ \left< \frac{1}{r} \sum_{\vec{k}'} \left( i m' n_{\vec{k}'} \pdr \phi_{\vec{k}-\vec{k}'} n_{\vec{k}}^*  - i (m - m') \pdr n_{\vec{k}'} \phi_{\vec{k}-\vec{k}'} n_{\vec{k}}^*        \right) \right>.
\eeqar

Finally, taking the real part of this equation results in:

\beqar
\label{real_density_evolution}
\left< \frac{1}{2} \pdiff{|n_{\vec{k}}|^2}{t} \right> = Re \left\{ \left< S_{N k_z} n_{m=0,k_z}^* \right> \right\} \nonumber \\
Re \left\{ \left< -\frac{i m}{r} \pdr N_0 \phi_{\vec{k}} n_{\vec{k}}^* - i k_z N_0 v_{\vec{k}} n_{\vec{k}}^* + \mu_N( \pdrr n_{\vec{k}} + \frac{1}{r} \pdr n_{\vec{k}} - \frac{m^2}{r^2} n_{\vec{k}}) n_{\vec{k}}^* \right> \right\} \nonumber \\
+ Re \left\{ \left< \frac{1}{r} \sum_{\vec{k}'} \left( i m' n_{\vec{k}'} \pdr \phi_{\vec{k}-\vec{k}'} n_{\vec{k}}^*  - i (m - m') \pdr n_{\vec{k}'} \phi_{\vec{k}-\vec{k}'} n_{\vec{k}}^*        \right) \right> \right\}.
\eeqar

Note that taking the real part of the equation produces the expected energy-like term on the left hand side because:

\beq
\frac{1}{2} \pdiff{|n_{\vec{k}}|^2}{t} = Re \left\{ \pdiff{n_{\vec{k}}}{t} n_{\vec{k}}^* \right\}.
\eeq

Breaking the result into explicit parts:

\beqar
\label{Fourier_density_evolution}
\pdiff{E_n(\vec{k})}{t} & = & Q_n(\vec{k}) + C_n(\vec{k}) + D_n(\vec{k}) + \sum_{\vec{k}'} T_n(\vec{k},\vec{k}') \\
E_n(\vec{k}) & = & \frac{1}{2} \left< |n_{\vec{k}}|^2 \right> \\
Q_n(\vec{k}) & = & Re \left\{ \left< -\frac{i m}{r} \pdr N_0 \phi_{\vec{k}} n_{\vec{k}}^* \right> \right\} \\
C_n(\vec{k}) & = & Re \left\{ \left< - i k_z N_0 v_{\vec{k}} n_{\vec{k}}^* \right> \right\} \\
D_n(\vec{k}) & = & Re \left\{ \left<  \mu_N( \pdrr n_{\vec{k}} + \frac{1}{r} \pdr n_{\vec{k}} - \frac{m^2}{r^2} n_{\vec{k}}) n_{\vec{k}}^*  + S_{N k_z} n_{m=0,k_z}^*  \right> \right\} \\
T_n(\vec{k},\vec{k}') & = & Re \left\{ \left< \frac{1}{r} \left( i m' n_{\vec{k}'} \pdr \phi_{\vec{k}-\vec{k}'} n_{\vec{k}}^*  - i (m - m') \pdr n_{\vec{k}'} \phi_{\vec{k}-\vec{k}'} n_{\vec{k}}^*        \right) \right> \right\}
\eeqar

$Q_n(\vec{k})$ is the energy injection, $C_n(\vec{k})$ is the transfer channel, $D_n(\vec{k})$ is dissipation, and $T_n(\vec{k},\vec{k}')$ is spectral energy transfer.
The same type of procedure may be applied to Eqs.~\ref{ve_eq}-\ref{te_eq}. 
However, the double primed conjugate multiplications (as in the step between Eqs.~\ref{density_eq_fourier} and~\ref{density_evolution}) 
must be done with the Fourier fields, $\fmei v_{\vec{k}''}$,  $- \phi_{\vec{k}''}$, and $\frac{3}{2} t_{\vec{k}''}$ rather than 
$v_{\vec{k}''}$, $\varpi_{\vec{k}''}$, and $t_{\vec{k}''}$.
These produce the correct energy terms, and most importantly still conserve the nonlinearities. The corresponding expressions for the perpendicular kinetic energy are:

\beqar
\label{Fourier_phi_evolution}
\pdiff{E_\phi(\vec{k})}{t} & = & Q_\phi(\vec{k}) + C_\phi(\vec{k}) + D_\phi(\vec{k}) + \sum_{\vec{k}'} T_\phi(\vec{k},\vec{k}') \\
E_\phi(\vec{k}) & = & \frac{1}{2} \left<  N_0 \left| \pdiff{\phi_{\vec{k}}}{r} \right|^2 + N_0 \frac{m^2}{r^2} |\phi_{\vec{k}}|^2  \right>\\
Q_\phi(\vec{k}) & = & 0 \\
C_\phi(\vec{k}) & = & Re \left\{ \left< i k_z N_0 v_{\vec{k}} \phi_{\vec{k}}^* \right> \right\} \\
D_\phi(\vec{k}) & = & Re \left\{ \left<  - \mu_\phi( \pdrr \varpi_{\vec{k}} + \frac{1}{r} \pdr \varpi_{\vec{k}} - \frac{m^2}{r^2} \varpi_{\vec{k}}) \phi_{\vec{k}}^* -  \nuin E_\phi(\vec{k})\right> \right\} \\
T_\phi(\vec{k},\vec{k}') & = & Re \left\{ \left< - \frac{1}{r} \left( i m' \varpi_{\vec{k}'} \pdr \phi_{\vec{k}-\vec{k}'} \phi_{\vec{k}}^*  - i (m - m') \pdr \varpi_{\vec{k}'} \phi_{\vec{k}-\vec{k}'} \phi_{\vec{k}}^*        \right) \right> \right\}
\eeqar

and for the electron temperature potential energy:

\beqar
\label{Fourier_te_evolution}
\pdiff{E_t(\vec{k})}{t} & = & Q_t(\vec{k}) + C_t(\vec{k}) + D_t(\vec{k}) + \sum_{\vec{k}'} T_t(\vec{k},\vec{k}') \\
E_t(\vec{k}) & = & \frac{3}{4} \left< |t_{\vec{k}}|^2  \right> \\
Q_t(\vec{k}) & = & Re \left\{ \left< - \frac{3}{2} \frac{i m}{r} \pdr T_{e0} \phi_{\vec{k}} t_{\vec{k}}^* \right> \right\} \\
C_t(\vec{k}) & = & Re \left\{ \left<  - 1.71 i k_z T_{e0} v_{\vec{k}} t_{\vec{k}}^* \right> \right\} \\
D_t(\vec{k}) & = & Re \left\{ \left< -\frac{\kpe}{N_0} k_z^2 |t_{\vec{k}}|^2  - \frac{3 m_e}{m_i} \nue |t_{\vec{k}}|^2 
+ \frac{3}{2} \mu_T( \pdrr t_{\vec{k}} + \frac{1}{r} \pdr t_{\vec{k}} - \frac{m^2}{r^2} t_{\vec{k}}) t_{\vec{k}}^*  + \frac{3}{2} S_{T k_z} t_{m=0,k_z}^*  \right> \right\} \\
T_t(\vec{k},\vec{k}') & = & Re \left\{ \left< \frac{3}{2 r} \left( i m' t_{\vec{k}'} \pdr \phi_{\vec{k}-\vec{k}'} t_{\vec{k}}^*  - i (m - m') \pdr t_{\vec{k}'} \phi_{\vec{k}-\vec{k}'} t_{\vec{k}}^*        \right) \right> \right\}
\eeqar

and for the parallel kinetic energy:

\beqar
\label{Fourier_vpar_evolution}
\pdiff{E_v(\vec{k})}{t} & = & Q_v(\vec{k}) + C_v(\vec{k}) + D_v(\vec{k}) + \sum_{\vec{k}'} T_v(\vec{k},\vec{k}') \\
E_v(\vec{k}) & = & \frac{1}{2} \fmei \left< |v_{\vec{k}}|^2 \right> \\
Q_v(\vec{k}) & = & Re \left\{ \left<  i k_z \frac{N_0^2 - T_{e0}}{N_0} n_{\vec{k}} v_{\vec{k}}^* + i k_z (1 - N_0) \phi_{\vec{k}} v_{\vec{k}}^* + 1.71 i k_z (T_{e0} -1) t_{\vec{k}} v_{\vec{k}}^*   \right> \right\} \\
C_v(\vec{k}) & = & Re \left\{ \left< - i k_z N_0 n_{\vec{k}} v_{\vec{k}}^* + i k_z N_0 \phi_{\vec{k}} v_{\vec{k}}^* - 1.71 i k_z T_{e0} t_{\vec{k}} v_{\vec{k}}^*  \right> \right\} \\
D_v(\vec{k}) & = & Re \left\{ \left< - \nue \fmei |v_{\vec{k}}|^2   \right> \right\} \\
T_v(\vec{k},\vec{k}') & = & Re \left\{ \fmei \left< \frac{1}{r} \left( i m' v_{\vec{k}'} \pdr \phi_{\vec{k}-\vec{k}'} v_{\vec{k}}^*  - i (m - m') \pdr v_{\vec{k}'} \phi_{\vec{k}-\vec{k}'} v_{\vec{k}}^*        \right) \right> \right\}
\eeqar

The transfer channel $C_v(\vec{k})$ is specifically set so that $C_n(\vec{k}) + C_t(\vec{k}) + C_\phi(\vec{k}) + C_v(\vec{k}) = 0$. 
The source $Q_v(\vec{k})$ is the left over quantity, which can have any sign and contributes to the overall energy evolution.



%
%

\end{document}